\documentclass[prb, twocolumn, dvipdfmx]{revtex4-1}
\usepackage{times}
\usepackage{amsmath, amssymb, bm}
\usepackage{graphicx, color, colortbl}
\graphicspath{{/Users/Yo/Library/Mobile\ Documents/rsc/real_spin_texture/}, {/Users/Yo/Pictures/}}
\usepackage{subfigure}
\usepackage{here}
\usepackage{url}
\begin{document}
\title{Large anomalous Nernst coefficient in an oxide Skyrmion crystal Chern insulator}
\date{\today}
\author{Yo Pierre Mizuta}
\author{Hikaru Sawahata}
\affiliation{Graduate School of Natural Science and Technology, Kanazawa University} 
\author{Fumiyuki Ishii}
\email{ishii@cphys.s.kanazawa-u.ac.jp}
\affiliation{Faculty of Mathematics and Physics, Kanazawa University, Kakuma, Kanazawa, Ishikawa 920-1102} 
\begin{abstract}
A sizable transverse thermoelectric coefficient $N$, large to the extent that it potentially serves applications, is predicted to arise, by means of first-principles calculations, in a Skyrmion crystal assumed on EuO monolayer where carrier electrons are introduced upon a quantum anomalous Hall insulating phase of Chern number $\mathcal{C}=2$. This encourages future experiments to pursue such an effect.
\end{abstract}
\maketitle
\section{Introduction}
Thermoelectric(TE) conversion, i.e. generation of electric power from heat, offers us a way of reviving enormous amount of waste heat, paving the way for a more energy-efficient society. Its wide applications, however, are hindered by various factors, among which the most basic one is the still limited performance of materials presently available. It is strongly hoped that better materials characterized by a larger value of the figure of merit $Z_XT$ will be found. Herein defined is $Z_X\equiv\sigma X^2 /\kappa$, where $\sigma$ and $\kappa$ are the longitudinal electrical and thermal conductivity respectively, $T$ is the mean temperature, and $X$ stands for Seebeck (Nernst) coefficient $S$ ($N$) quantifying the generated electric field ${\bf E}$ parallel (perpendicular) to the temperature gradient $\nabla T$ [Seebeck(Nernst) voltage]. While the Seebeck-based TE modules have been at the center of applied studies, the Nernst-based ones have various advantages over the former: simpler and more flexible structure that facilitates its fabrication and its application to heat sources with non-flat surfaces. The minimum $|N|$ for the latter to be useful was estimated to be $\sim 20\mu {\rm V/K}$. \cite{Sakuraba_Potential_2016} 

Non-zero $N$ can appear in a material only when at least either of the transverse currents $j_x^{\rm (H)} = \sigma_{xy}E_y$ or $j_x^{\rm (N)}=\alpha_{xy}\nabla_yT$ is present, whose conductivities relate to the TE coefficients as described in Sec.II. It can be shown for a two-dimentional system on the $xy$ plane, that this condition requires the breaking of both the time-reversal symmetry (TRS) and the mirror symmetry (MS) with respect to planes normal to the $xy$ plane, which is realized obviously due to an externally applied \textit{real} magnetic field $B_z^{\rm ext}$, but also due to a field $B_z^{\rm int}$ that effectively arises from internal degree(s) of freedom of the material and posesses the same symmetry property as that of $B_z^{\rm ext}$. The $B_z^{\rm int}$ in a crystalline material is generally identified to be the \textit{Berry curvature}\cite{Xiao_Berry_2010} $\boldsymbol{\Omega}(\bm{k})\equiv i \langle \partial_{\bm{k}}u |\times |\partial_{\bm{k}}u \rangle$, defined in the space of the lattice periodic part of Bloch states $|\psi_{\bm{k}} \rangle = e^{i\bm{k}\cdot \hat{\bm{r}}}|u_{\bm{k}} \rangle$ parameterized with crystal momentum $\bm{k}$. In the presence of $B_z^{\rm ext}$($B_z^{\rm int}$), the consequent emergence of $j_x^{\rm (H)}$ and $j_x^{\rm (N)}$ are called the ordinary(anomalous) Hall effect [OHE(AHE)] and the ordinary(anomalous) Nernst effect [ONE(ANE)], respectively. In terms of applications, the use of OHE and ONE, which involves the attachment of probably strong magnets to the TE converter, is not practically convenient, while the use of AHE and ANE is not yet realistic with the small values $|N| \lesssim 1 \mu{\rm V/K}$ of most of the materials so far investigated. Therefore the question of particular interest is: \textit{How large $|N|$ could be, by virtue of AHE and ANE? } 

The necessary condition to obtain AHE and/or ANE in a given system, 
 which is the \textit{lack of TRS and MS} without $B_z^{\rm ext}$ as stated above, can be satisfied through either of the following two scenarios: (i) TRS breaking due to a \textit{coplanar} magnetic (CM) structure, combined with MS breaking due to spin-orbit interaction (SOI)\cite{Chen_Anomalous_2014, Ikhlas_Large_2017} and (ii) simultaneous breaking of TRS and MS due to a \textit{non-coplanar} magnetic (NCM) structure.\cite{Zhou_Predicted_2016}\footnote{Several mechanisms are known to generate NCM structures: Dzyaloshinskii-Moriya interaction, which is a particular form of SOI arising in systems without inversion center, or the electron-electron Coulomb interaction in a system with trigonal or hexagonal lattice symmetry, where it effectively creates frustrated exchange interactions among spins\cite{Batista_Frustration_2016, Okubo_Multiple_2012}.} 

In this paper, we concentrate on (ii), and out of many possible NCM patterns, particularly on the {\it Skyrmion}, which has received special interests associated with its quantized (particle-like) nature,\cite{Nagaosa_Topological_2013} and here exclusively consider its crystalline form \textit{Skyrmion crystal}(SkX), on which some experimental studies on the ANE also exist\cite{Shiomi_Topological_2013, Hirokane_Longitudinal_2016} as well as many other studies on AHE.
Following our previous work\cite{Mizuta_Large_2016} predicting a surprisingly large ANE in a similar model of SkX whose AHE had been investigated by Hamamoto {\it et al.},\cite{Hamamoto_Quantized_2015} we now target a more realistic system by taking advantage of large scale first-principles calculations. 
Namely we pick EuO as a promising material, which is a rare ferromagnetic insulator as a bulk with half-filled 4$f$-shells forming Heisenberg-type spins, and whose thin films indicated a Skyrmion-like magnetic structure in a recent experiment.\cite{Ohuchi_Topological_2015}
The results show that SkX is indeed such a NCM structure that can support large $|N|$.  

This paper is organized as follows: In Sec.II, we present the expression of $N$ (and also $S$) evaluated in this study, at the same time noting the simple relation between AHE and ANE, which is very instructive in any attempt on larger $|N|$. In Sec.III, we describe the details of our EuO model, both its geometry and the choice of parameters. In Sec.IV, we summarize the employed computational procedure based on first-principles methods. Sec.V is the main part giving results and discussions on the electronic structure and the associated transport coefficients. The key finding was that, our model becomes a \textit{Chern insulator}\footnote{A class of topologically non-trivial phase, characterized by Chern number $\mathcal{C}$, which is a topological invariant and determines a quantized conductivity of AHE as $\sigma_{xy} = \mathcal{C}(e^2/h)$.} 
of \textit{Chern number} $\mathcal{C}=2$, doping of some carriers on which leads to large $|N|$ due to the coexistence of large Seebeck effect and and large AHE.
It is hoped that our results to be reported here will give us positive prospects for finding large $|N|$ in real, multi-orbital systems.
\section{Expressions of TE quantities}
The formulae for the thermoelectric cofficients to be evaluated follow from the linear response relation of charge current: ${\bf j}=\tilde{\sigma}{\bf E}+\tilde{\alpha}(-\nabla T)$.  
Using the conductivity tensors $\tilde{\sigma}=[\sigma_{ij}]$ and $\tilde{\alpha}=[\alpha_{ij}]$, we obtain \cite{Mizuta_Large_2016}
\begin{equation} \left\{
 \begin{array}{l}
{\displaystyle S\equiv S_{ii} \equiv \frac{E_i}{\nabla_i T}=\frac{S_0}{1+r_H^2} + \frac{r_H N_0}{1+r_H^2}}
 \vspace{2mm}\ \\
{\displaystyle N \equiv S_{xy} \equiv \frac{E_x}{\nabla_y
 T}=\frac{N_0}{1+r_H^2} - \frac{r_H S_0}{1+ r_H^2}} \label{SN}
\end{array} \right.
\end{equation}
Here we defined $S_0 \equiv \alpha_{xx}/\sigma_{xx}$ (\textit{pure Seebeck coefficient}), $r_H \equiv \sigma_{xy}/\sigma_{xx}$  ({\it Hall angle ratio}), $N_0 \equiv \alpha_{xy}/\sigma_{xx}$ (\textit{pure Nernst coefficient}) for a simpler notation.
The first term in each of $S$ and $N$, which is proportional to pure coefficient $S_0$ and $N_0$ respectively, survives in the vanishing AHE condition $r_H=0$, while the second term are present only when $r_H \ne 0$. 
The sign relation among $S_0$, $N_0$ and $r_H$ determines whether the two terms work constructively or destructively in each of the resultant $S$ and $N$, as will be summarized later in TABLE I.
In the presence of AHE and ANE ($r_H \ne 0,\ N_0 \ne 0$), the mixture of pure coefficients $S_0$ and $N_0$ gives the measured coefficients $S$ and $N$.

Importantly, the AHE and ANE are related via simple formula,\footnote{See for example, the description between Eq.(1) and (2) of Ref.\onlinecite{Mizuta_Large_2016}} which, at low temperatures in particular, reduces to the well-known Mott relation:
\begin{align}
\alpha_{ij}(T,\ \mu \simeq \varepsilon_F) &\simeq \frac{\pi^2k_{\rm B}^2}{3e}\frac{d\sigma_{ij}(0, \varepsilon_F)}{d\varepsilon_F}T,  \label{mott_lowT}
\end{align}
which instructs us to seek for $\sigma_{ij}(\varepsilon)$ that varies more rapidly at $\varepsilon_F$, i.e. larger Berry curvature at $\varepsilon_F$, in order to achive larger $N_0$.  

\section{Model}
\begin{figure}
	\begin{center}
		\includegraphics[width=0.50\textwidth]{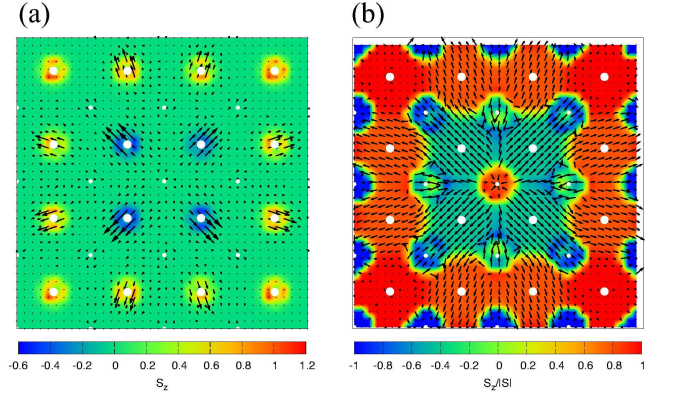}
	\end{center}
	\caption{(a) Spin density $\bm{S}(\bm{r})$ evaluated from the wave functions on grid points. The out-of-plane component $S_z$ is expressed with color, while the in-plane components are represented with the black arrows. (b) A plot corresponding to the left one, but for normalized spin density $\bm{S}(\bm{r})/|\bm{S}(\bm{r})|$. Both panels illustrate a single square unitcell of $4\times4$ SkX, with the length of each side being $2\lambda$. The Eu (O) atoms are indicated with large (small) white filled circles.\label{skx} }
\end{figure}

We choose, as a realistic material, EuO, which is a rare ferromagnetic insulator in bulk with half-filled 4$f$-shells forming Heisenberg-type spins.
We specifically consider its monolayer (two-dimensional limit) for ease of comparison with our previous report \cite{Mizuta_Large_2016}. A 2D square unitcell is constructed as a $2\sqrt{2}\times 2\sqrt{2}$ cut with respect to the cell vectors of FCC lattice, to which the rocksalt structure of bulk EuO belongs. The lattice constant was set to $2\lambda \equiv (a=5.14)\times 2\sqrt{2}=14.54$ \AA, adopting the bulk value as $a$.\cite{Wachter_Optical_1972} A unitcell contains 16 Eu atoms whose spins we force to form a $4\times4$ skyrmion structure equivalent to the one previously studied\cite{Hamamoto_Quantized_2015}, i.e., the spin spherical coordinates are set as $\theta(\bm{r})=\pi(1-r/\lambda)$ for $r<\lambda$ and $\theta=0$ for $r>\lambda$, along with $\phi=\tan^{-1}(y/x)$. An image of unitcell is shown in Fig.\ref{skx}.
%
%
The noncoplanar \{$\theta(\bm{r}),\ \phi(\bm{r})$\} gives rise to an emergent magnetic field $B_{\rm spin}(\bm{r})=(\hbar/2e)(\pi/r\lambda)\sin \pi(1-r/\lambda)$,\footnote{\textcolor{black}{The general description with $\boldsymbol{\Omega}(\bm{k})$ in $\bm{k}$-space can be translated into real space picture, in the approximation of strong Hund's coupling, i.e. frozen spin, where the ``spin-less" electrons feel ``real" magnetic field $B_{\rm spin}$\cite{Hamamoto_Quantized_2015, Gbel_Unconventional_2017}.}} whose spatial average in a unitcell is $\overline{B}_{\rm spin}=(h/e)/(\pi \lambda^2)$. Thus, the associated cyclotron frequency in our model is evaluated to be $\overline{\omega_s}=e\overline{B}_{\rm spin}/m^* \approx 10^{14}\ {\rm s}^{-1}$, given the effective mass of \textcolor{black}{$m^{*} \approx 0.3$ reported by Ref.\onlinecite{Schoenes_Exchange_1974}.}    
As to the scattering effects on electrons, we adopt the constant-relaxation-time($\tau$) approximation. While $\tau=m^* \mu /e \approx 100$ fs is roughly estimated from a mobility of $\mu \approx 30$ cm$^2$V$^{-1}$s$^{-1}$ experimentally reported,\cite{Ohuchi_Topological_2015} we will consider three different values of $\tau=10\ {\rm fs}, \ 100\ {\rm fs},\ 1\ {\rm ps}$, since we can expect $\tau$ to vary substantially depending on the sample details. Correspondingly, electrons can typically complete as many cycles as $(\overline{\omega_s}\tau)/(2\pi)$ times between successive scattering events, with $\overline{\omega_s}\tau \approx 1,\ 10,\ 100$. Therefore, especially the choice of larger two values justifies our picking up only the purely intrinsic Berry-curvature-driven term in $\sigma_{xy}$ and $\alpha_{xy}$,  which is exact for $\overline{\omega_s}\tau \gg 1$.

\section{Computational procedure}
Our calculations consist of three steps: (1)Obtaining the electronic states of target SkX using {\it OpenMX},\cite{openmx} (2)Constructing Wannier functions employing {\it Wannier90}\cite{Mostofi_In_2014}  and finally (3) Computing all the necessary transport quantities[tensors $\sigma$ and $\alpha$ in Eq.(\ref{SN})] in the obtained Wannier basis using a postprocessing code {\it postw90}.\cite{Pizzi_BoltzWann_2014} 

In step (1), the calculation was performed based on the density functional theory within Perdew-Burke-Ernzerhof's generalized gradient approximation  of exchange-correlation energies, including a Hubbard $U$ correction of $U=6.0$ eV for the localized Eu-$f$ orbitals. Norm-conserving pseudopotentials were used, and a set of pseudoatomic orbital basis was specified as Eu8.0-$s$2$p$2$d$2$f$1 and O5.0-$s$2$p$2$d$1, where the number after each element standing for the radial cutoff in the unit of bohr, and the integer after $s,p,d,f$ indicating the radial multiplicity of each angular momentum component. Cutoff energy of 300 Ry for the charge density, and a $k$-points grid of  $4\times4\times 1$ were adopted.

In step(2), from a total of 214 bands falling inside the energy window of [0, 10]eV, 192 Wannier orbitals were extracted starting from projecting the original Bloch states sampled on a $k$ grid of $4\times4\times 1$ to a set of ($s,\ d$)-character localized orbitals on Eu sites. The states inside [0, 3]eV were frozen during the extraction process, so that the original band dispersion was maintained as is required for the exact evaluation of $\sigma_{xx}$ and $\alpha_{xx}$ and necessary for a reasonable evaluation of $\sigma_{xy}$ and $\alpha_{xy}$.   

In step(3), all the conductivities were evaluated within the constant-relaxation-time approximation to the semiclassical Boltzmann transport theory,\footnote{See for example Eq.(2) of Ref.\onlinecite{Mizuta_Large_2016} for the evaluated mathematical expressions.} where the numerical integrations in $k$ space were performed on a grid of $100\times100\times1$.

In addition to the above energy-decomposed analysis, we have also performed band-by-band identification of Chern number $C_n$ ($n$: band index), in the aim of clarifying the topological structure of the electronic bands.

\section{Results and Discussions}
\subsection{Basic results}
\textcolor{black}{Let us first look at the landscape of spin momentum density obtained after the self-consistent calculation. As shown in Fig.\ref{skx}(a), large momentum is strongly localized around Eu atoms, which justifies the ionic picture of Eu-$4f$ shell. Furthermore, its directional distrubution in the vicinity of Eu atoms is scarecely changed from the initial setting of our Skyrmion texture as described in the previous section, suggesting the spin-constraining method in {\it OpenMX} worked effectively during the self-consistent process.
It is also interesting to pay attention to the directional distribution in the small-moment region: The normalized spin density vector drawn in Fig.\ref{skx}(b) clearly demonstrates a strongly antiferromagnetic coupling between Eu and O atoms, although the latter have much smaller($<0.2\mu_B$) momentum. The effect of the observed fine structure on the transport properties of our interest is beyond the scope of this paper.} 

\begin{figure}
	\begin{center}
		\includegraphics[width=0.50\textwidth]{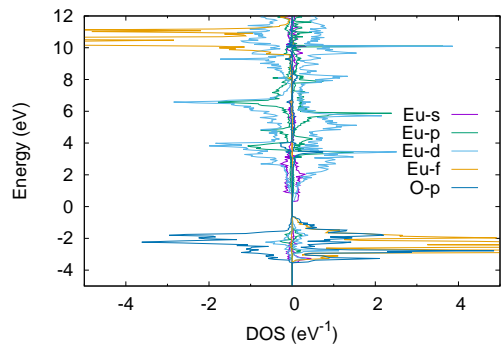}
	\end{center}
	\caption{Spin-, atom- projected density of states of collinear ferromagnetic EuO monolayer in a broad range of energy.\label{dos}}
\end{figure}

\textcolor{black}{Before discussing the electronic structure of the SkX, we look at the density of states(DOS) of collinear ferromagnetic (FM) state in a broad range of energy as shown in Fig.\ref{dos}.  It was confirmed to be reasonable with the choice of $U=6.0$ eV, in the sense that the position of the Eu-$f$ levels, which is roughly 2 eV below the conduction band bottom are consistent with what had been obtained from the all-electron method for bulk EuO.\cite{Miyazaki_Direct_2009} Perfect polarization of Eu-$f$ spins ($\sim 7\mu_B$) is indicated, while those states are absent in the range of ~10eV from the conduction band bottom, where Eu-$s$, -$p$, -$d$ states are dominant.} 
\begin{figure}
	\begin{center}
		\includegraphics[width=0.50\textwidth]{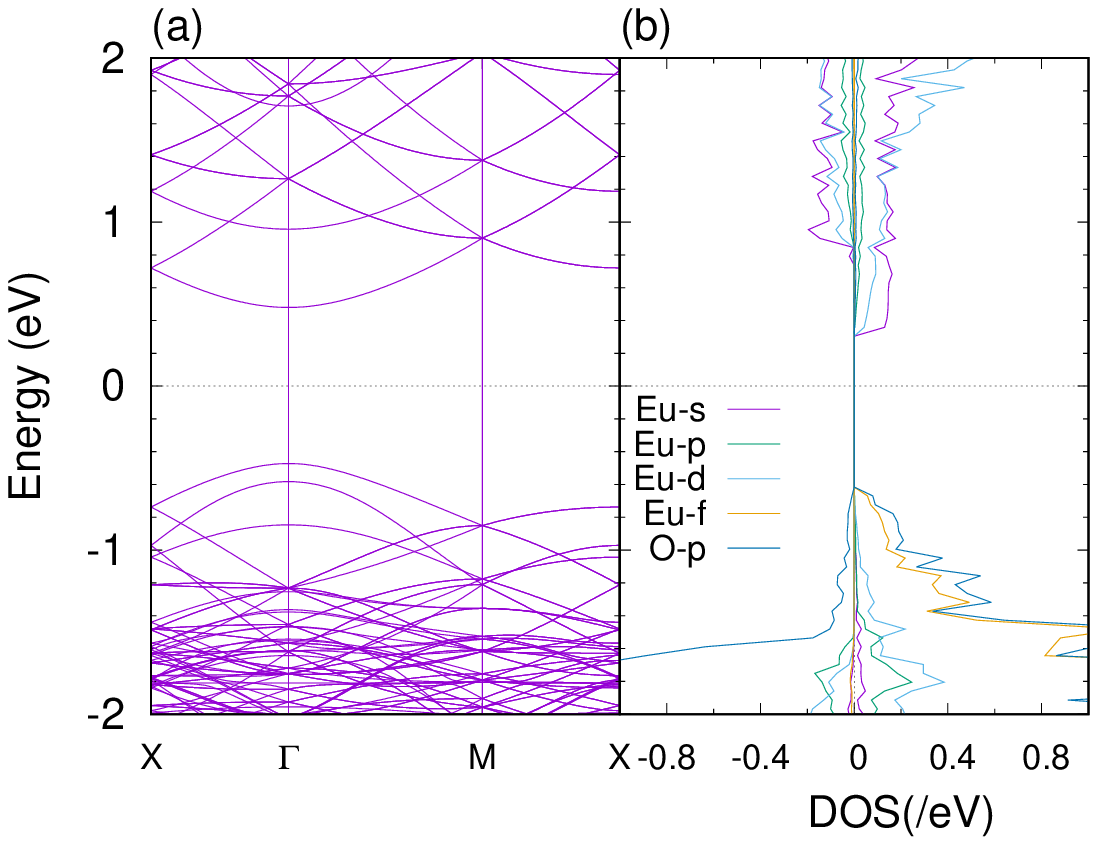}
	\end{center}
	\caption{(a)Band structure and the (b)spin-projected density of states for the FM system, focused around the Fermi energy.\label{b+dos_zoom_FM}}
\end{figure}
\begin{figure}
	\begin{center}
		\includegraphics[width=0.50\textwidth]{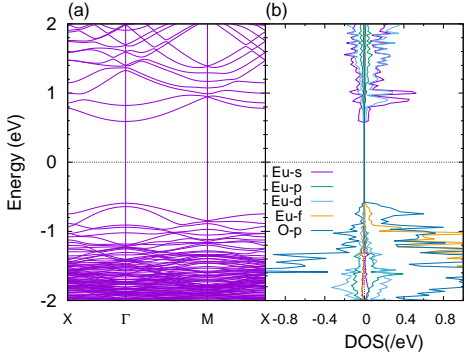} 
	\end{center}
	\caption{(a)Band structure and (b)the spin-projected density of states for the SkX system, focused around the Fermi energy.\label{b+dos_zoom_skx}}
\end{figure}
\subsection{Effects of SkX on electronic structure}
Hereafter we focus on the lower energy region of Fig.\ref{dos}, whose enlarged DOS and band dispersion are shown in Fig.\ref{b+dos_zoom_FM}.  
The corresponding plots for the SkX system are in Fig.\ref{b+dos_zoom_skx}. 
While bands are degenerate on symmetry points or lines in FM case, they become gapped once the spins form SkX. This can be understood as a consequence of symmetry lowering due to the appearance of emergent magnetic field $B_z^{\rm int}$, with the unit length of translation enlarged from that in the FM state. 
Consequently, the bands become narrower, exhibiting many local gaps (anti-crossings), and accordingly the DOS shows more ups-and-downs in SkX than in FM. Particularly, there is a global(equi-energy) gap of $\sim$20 meV between the second and the third band (counted from the conduction band bottom). This band structure is a realistic example of a Landau level split and dispersed by the crystal potential, so far reported in tight-binding models under a uniformly applied magnetic field\cite{Arai_Quantum_2009} or a SkX-driven inhomogeneous emergent field, \cite{Hamamoto_Quantized_2015, Gbel_Unconventional_2017} the latter being the same phisical situation as ours. 

\begin{figure}
	\begin{center}
		\includegraphics[width=0.5\textwidth]{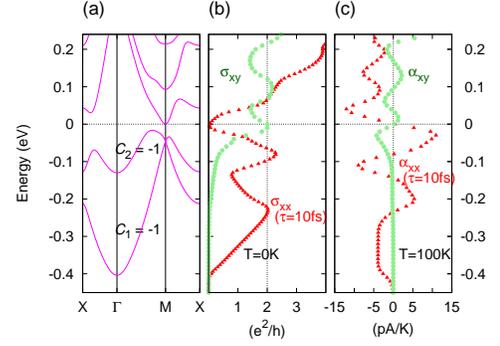}
	\end{center}
	\caption{(a) Zoomed-in band structure around the conduction band bottom, and the chemical potential dependence of (b) zero-temperature electric conductivities $\sigma_{xx}(\tau=10\ {\rm fs})$ and $\sigma_{xy}$, and of \textcolor{black}{(c) thermoelectric conductivities $\alpha_{xx}(\tau=10\ {\rm fs})$ and $\alpha_{xy}$ at 100 K.} The system has two additional electrons as compared to the pristine system, which occupy the lowest two bands in (a), where their respective Chern numbers of -1 are indicated. A Gaussian smearing of 10 meV is employed for (b).  \label{b+sig}}
\end{figure}

A clear evidence of an emergent field in our system is the non-zero Chern numbers we obtained, for example $C_1=C_2=-1$ (labeling the conduction bands as 1, 2, ... from the bottom)[Fig.\ref{b+sig}]. This is consistent with the earlier observations, based on Onsager's semiclassical-quantization argument, that the Chern number roughly correponds to the number of topologically disconnected Fermi surfaces of the original bands in the absence of magnetic field\cite{Arai_Quantum_2009, Gbel_Unconventional_2017}: In fact, in the energy range of $n=1, 2$ bands, there is only one electron pocket around $\Gamma$ point in the absence of SkX structure, i.e., when the bands of FM EuO shown in Fig.\ref{b+dos_zoom_FM}(a) are unfolded to the primitive Brillouin zone.     

\textcolor{black}{An important finding here is that our SkX becomes a Chern insulator of $\mathcal{C}=2$ when it is doped with two additional electrons per SkX unitcell(0.125 electron per Eu atom), filling the bands $n=1,\ 2$. This had already been implied in the pristine system (Fig.\ref{b+dos_zoom_skx}), and indeed confirmed by an explicit self-consistent calculation including two additional electrons (Fig.\ref{b+sig}).
With the Fermi energy located inside the previouly mentioned global gap of $\sim 20$ meV, the Hall conductivity is quantized to be $\sigma_{xy} = 2(e^2/h)$ in the insulating phase($\sigma_{xx}=0$)[Fig.\ref{b+sig}(b)].}

\textcolor{black}{From a categorizing point of view, our SkX is an example of Chern insulators driven by an emergent magnetic field $B_z^{\rm int}$ arising from \textit{Hund's coupling} of the conduction electrons to the SkX background, while a series of EuO/GdN hetero films predicted to be Chern insulators\cite{Garrity_Chern_2014} were driven by $B_z^{\rm int}$ arising from \textit{spin-orbit coupling}.} 

\subsection{Thermoelectric response largely modulated by SkX}
\begin{figure}
	\begin{center}
	\includegraphics[width=0.50\textwidth]{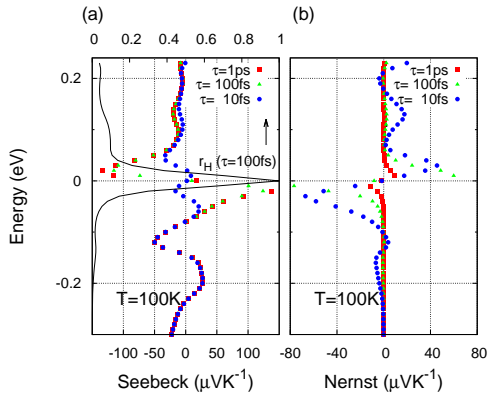}
	\end{center}
	\caption{Chemical potential dependence of TE coefficients: (a)$S$ and (b)$N$ at 100 K, respectively. Results are shown for three different values of $\tau$ in each case. Additionally, the Hall angle ratio $r_H$ for $\tau=100$ fs is plotted on the second axis in (a) [black solid line]. \label{TE}} 
\end{figure}

\textcolor{black}{Now we will discuss our main subjects, the thermoelectric quantities of the system, focusing on a fixed temperature $T=100$ K, which is below the highest Curie temperatures so far reported, e.g. in La-doped EuO thin films.\cite{Miyazaki_La_2010} The computed Seebeck/Nernst coefficients are shown in Fig.\ref{TE} for three different strengths of disorder parametrized by a relaxation time $\tau=10\ {\rm fs}, \ 100\ {\rm fs},\ 1\ {\rm ps}$ of the distribution of electrons.  
Our principal result is the large $N$ expected for a moderate value of $\tau$=100 fs, reaching $N\approx 20\ \mu$V/K at $E_{\rm F}\simeq 30$ meV, which becomes an order of magnitude smaller for $\tau$=1 ps. This is in contrast to $|S|$, which is about five times larger than $N$ at that $E_{\rm F}$ for $\tau$=100 fs and remains almost the same for the ten times longer $\tau$.
To understand such behavior of $S$ and $N$ in terms of their three constituents of distinct origins, namely $S_0,\ N_0$ and $r_H$, we pay attention to the sign relation among the three and the magnitude of $r_H$. In the following we focus on around $\mu \sim \pm 30$ meV,\footnote{\textcolor{black}{For the semiclassical picture of band conduction to be valid, $E_{\rm F}\tau \gg 1$ is required, and therefore we should limit ourselves to $E_{\rm F} \gg 50\ (5)$meV for $\tau=10\ (100)$ fs. To discuss as large $N$ as possible but safely enough, we specifically discuss $\mu \sim \pm 30$ meV.}} as well as on the choice of $\tau \ge 100$ fs for the reason stated at the end of Sec.III.}

\textcolor{black}{The sign of $\sigma_{ij}$ and $\alpha_{ij}$ can be read from Fig.\ref{b+sig}(b) and (c).}
The two patterns we find in the range $|\mu| \lesssim 30$ meV are listed in TABLE I.
Note that, not only the sign patterns but also the overall behavior of $\alpha_{ij}$ at $T=100$ K in (c), which is apparently well proportional to the derivative of $\sigma_{ij}|_{T=0}$ in (b), is consistent with the Mott relation, i.e. Eq.(\ref{mott_lowT}).
\begin{table}[h] 
 \caption{Sign of each quantity, and the relation between pure and measured coefficients. Two cases appearing in the vicinity of $\mu=0$ in Fig.\ref{TE} are picked.} 
 \begin{tabular}{|c||c|c|c||c|c|} \hline
  & $S_0$ & $N_0$ & $r_H$ & $|S|$ & $|N|$ \\ \hline 
  $\mu > 0$ & $-$ & $+$ & $+$ & $<|S_0|$ & $>|N_0|$ \\ \hline 
  $\mu < 0$ & $+$ & $-$ & $+$ & $<|S_0|$ & $>|N_0|$ \\ \hline
  \end{tabular}
\end{table}

 \textcolor{black}{As to the Hall angle ratio, we find from Fig.\ref{TE}(a) that $r_H(\mu=\pm 30\ {\rm meV}) \simeq 0.2$.
In this case, the expression Eq.(\ref{SN}) can be approximated up to the error of $\sim 5\%$ as $S \simeq S_0 + r_HN_0$ and $N \simeq N_0 - r_HS_0$.
They can be further approximated as} 
\begin{align}
S \simeq S_0,\  {\rm and}\  N \simeq -r_HS_0, \label{approxSN}
\end{align}
\textcolor{black}{
since the peaks in question satisfy $|S_0| \gtrsim (20 \sim 100)\times |N_0|$ [as understood by comparing the red points multiplied by a number $> 10$ (corresponding to $\tau \ge 100\ {\rm fs}$) and the green points in Fig.\ref{b+sig}(c)]. Therefore, despite the relation summarized in TABLE I, such a destructive (constructive) effect actually has little effect on $S$ ($N$) in the range of our focus, and importantly, the large $|N|$ is due to the combination of large Seebeck effect $\propto S_0$ and large AHE $\propto r_H$, as is clear from Eq.(\ref{approxSN}).}
    
\subsection{Prototypical cases for larger Nernst coefficient} 
\textcolor{black}{Although we obtained remarkably large $|N| \sim 20\ \mu{\rm V/K}$ in the above analysis, it is worth considering from wider perspective, so as to recognize different possibilities in other systems: Three prototypical cases (a-c) are listed in the right columns of TABLE II, where different quantities are responsible for large $|N|$. We particularly set $|N|\gtrsim 10\ \mu{\rm V/K}$ as a typical target,\footnote{\textcolor{black}{This is still not enough for wide applications, on which $|N| \sim 50\ \mu{\rm V/K}$ will have a significant impact.\cite{Sakuraba_Potential_2016}}} which is close to the record-high experimental value,\cite{Pu_Mott_2008} but an order of magnitude larger than the widely obtained experimental values $|N|<1\ \mu{\rm V/K}$. The (a-b) and (c) belong to the conditions (I) and (II), respectively.}
\textcolor{black}{An example of (c) is our previous finding, where possibility of large $|N|$ was proposed within a single $s$-orbital model, which is much simpler than the present EuO model. While the considered spin texture was a common SkX, the behavior of TE coefficients was essentially different: There we found a striking peak of $N$ around a specific band filling\footnote{\textcolor{black}{The band filling corresponds to $E_F$ close to a van Hove singularity(VHS). Drastical change of Chern numbers in the vicinity of VHS has already been observed\cite{Arai_Quantum_2009, Gbel_Unconventional_2017}, which received a novel interpretation recently\cite{Naumis_Topological_2016}.}} where the large voltage mostly originates from pure ANE, i.e. large $N_0$. On the other hand, the present case is rather close to (b), where large $|r_H|$ is essential ~\footnote{\textcolor{black}{The $|r_H| \sim 0.2$ is one order of magnitude larger than the typical largest value obtained in ``bad metal"(localization) regime. See for example the discussion in Supporting Online Material of Ref.\onlinecite{Lee_Dissipationless_2004}}.}.}
\textcolor{black}{In real materials, although conditions vary (I) through (II) depending on the temperatures or the chemical composition,\cite{Lee_Anomalous_2004} $|N|$ remains rather small $\sim 1\ \mu{\rm V/K}$, such that the situation is far from any of (a-c). It is therefore strongly hoped that the behavior close to any of (a-c) will be found in some materials. The case (c) seems particularly favorable in the sense that it can be a good thermoelectric system however small $S_0$ may be, i.e. materials that have never been considered to be useful could belong to (c).}
\begin{table}[h]
 \caption{Some limiting cases leading to large $N$ when $r_H^2 \ll 1$.} 
 \begin{tabular}{|c|c|c||c|c|c|c|} \cline{3-7}
\multicolumn{2}{c}{}			     & \multicolumn{5}{|c|}{Typical patterns for $N > 10\ \mu{\rm V/K}$}  			 	\\ \cline{1-6}
\multicolumn{2}{|c|}{Condition}		     &  $N$   		    	  & $|S_0|$		& $|r_H|$ 		& $|N_0|$ 	& 		\\ \hline
I  &  	    $|N_0| \ll r_HS_0$		     &  $\sim -r_HS_0$  	  & $> 10^3$ 	   	&   $\sim 10^{-2}$	& - 		&	(a)	  \\ \cline{4-7}
   &   				             & 				  & $> 10^2$ 		&   $\sim 10^{-1}$ 	& -		&	(b)	  \\ \hline
II &  	    $|N_0| \gtrsim r_HS_0$ 	     &  $\sim N_0$     		  &  -   		&        -		& $> 10$	&	(c)	  \\ \hline
  \end{tabular}
 \end{table}
\subsection{Connection to the experimental situation}
\textcolor{black}{Finally, we comment on the feasibility of experimentally achieving the situation close to Fig.\ref{b+sig} and Fig.\ref{TE}. The requirements are that (i) only about two conduction bands are occupied per Skyrmion unitcell and that (ii) $\omega_c\tau \gg 1$. These imply the following (i') and (ii'), respectively, for a SkX with unit Skyrmion that is assumed to be $x$ times larger (in linear dimension) than ours ($\sim$1 nm): (i') sheet electron density smaller at least by a factor of $x^{-2}$ than our $\approx 10^{14}\ {\rm cm}^{-2}$, and (ii') relaxation time $\tau$ longer at least by a factor of $x^2$ than our safe value 100 fs.  For example, the experimental results on LaAlO$_3$/SrTiO$_3$ heterostructures of Ref.\onlinecite{Xie_Quantum_2014} satisfy both requirements for $x \sim 10$. While this system itself shows magnetism,\cite{Brinkman_Magnetic_2007} systems of the same class but containing EuO layers, such as LaAlO$_3$/EuO \cite{Wang_Prediction_2009} or EuTiO$_3$\cite{Ahadi_Evidence_2017}, should better resemble our model and seem more promising, if carriers of similar quality, i.e. very dilute but mobile two-dimensional ones, can be induced.}
\section{Summary \& Conclusions}
\textcolor{black}{
 We showed from first-principles that, very large Nernst coefficient $|N| \sim 20\mu{\rm V/K}$ would appear in the presence of Skyrmion crystal magnetic order on a EuO monolayer. This is expected when a small number of carriers are introduced on top of the non-trivial insulating phase with Chern number -2, realized by the Skyrmion-driven emergent magnetic field. Experiments on very clean interfaces or thin films containing EuO layers would hopefully observe this effect. 
In view of future energy-saving applications, extensive studies are needed to realize such a giant Nernst effect in a wide range of more easily accessible systems hosting stable phases of Skyrmion crystal.}
\section*{Acknowledgements}
The authors thank N. Nagaosa and R. Arita for their insightful comments on this study.
This work was supported by JSPS KAKENHI Grant Number JP17J03672.
This work was partly supported by Grant-in-Aid for Scientific Research on Innovative Area, ``Nano Spin Conversion Science" (Grant Nos. 15H01015 and 17H05180). This work was also partly supported by Grants-in-Aid for Scientific Research (No.16K04875) from the Japan Society for the Promotion of Science.
The computations in this research were performed using the supercomputers
at ISSP, University of Tokyo.
\bibliography{readcube_export}

\begin{thebibliography}{40}%
\makeatletter
\providecommand \@ifxundefined [1]{%
 \@ifx{#1\undefined}
}%
\providecommand \@ifnum [1]{%
 \ifnum #1\expandafter \@firstoftwo
 \else \expandafter \@secondoftwo
 \fi
}%
\providecommand \@ifx [1]{%
 \ifx #1\expandafter \@firstoftwo
 \else \expandafter \@secondoftwo
 \fi
}%
\providecommand \natexlab [1]{#1}%
\providecommand \enquote  [1]{``#1''}%
\providecommand \bibnamefont  [1]{#1}%
\providecommand \bibfnamefont [1]{#1}%
\providecommand \citenamefont [1]{#1}%
\providecommand \href@noop [0]{\@secondoftwo}%
\providecommand \href [0]{\begingroup \@sanitize@url \@href}%
\providecommand \@href[1]{\@@startlink{#1}\@@href}%
\providecommand \@@href[1]{\endgroup#1\@@endlink}%
\providecommand \@sanitize@url [0]{\catcode `\\12\catcode `\$12\catcode
  `\&12\catcode `\#12\catcode `\^12\catcode `\_12\catcode `\%12\relax}%
\providecommand \@@startlink[1]{}%
\providecommand \@@endlink[0]{}%
\providecommand \url  [0]{\begingroup\@sanitize@url \@url }%
\providecommand \@url [1]{\endgroup\@href {#1}{\urlprefix }}%
\providecommand \urlprefix  [0]{URL }%
\providecommand \Eprint [0]{\href }%
\providecommand \doibase [0]{http://dx.doi.org/}%
\providecommand \selectlanguage [0]{\@gobble}%
\providecommand \bibinfo  [0]{\@secondoftwo}%
\providecommand \bibfield  [0]{\@secondoftwo}%
\providecommand \translation [1]{[#1]}%
\providecommand \BibitemOpen [0]{}%
\providecommand \bibitemStop [0]{}%
\providecommand \bibitemNoStop [0]{.\EOS\space}%
\providecommand \EOS [0]{\spacefactor3000\relax}%
\providecommand \BibitemShut  [1]{\csname bibitem#1\endcsname}%
\let\auto@bib@innerbib\@empty
\bibitem [{\citenamefont {Sakuraba}(2016)}]{Sakuraba_Potential_2016}%
  \BibitemOpen
  \bibfield  {author} {\bibinfo {author} {\bibfnamefont {Y.}~\bibnamefont
  {Sakuraba}},\ }\href {\doibase 10.1016/j.scriptamat.2015.04.034} {\bibfield
  {journal} {\bibinfo  {journal} {Scr. Mater.}\ }\textbf {\bibinfo {volume}
  {111}},\ \bibinfo {pages} {29} (\bibinfo {year} {2016})}\BibitemShut
  {NoStop}%
\bibitem [{\citenamefont {Xiao}\ \emph {et~al.}(2010)\citenamefont {Xiao},
  \citenamefont {Chang},\ and\ \citenamefont {Niu}}]{Xiao_Berry_2010}%
  \BibitemOpen
  \bibfield  {author} {\bibinfo {author} {\bibfnamefont {D.}~\bibnamefont
  {Xiao}}, \bibinfo {author} {\bibfnamefont {M.}~\bibnamefont {Chang}}, \ and\
  \bibinfo {author} {\bibfnamefont {Q.}~\bibnamefont {Niu}},\ }\href {\doibase
  10.1103/revmodphys.82.1959} {\bibfield  {journal} {\bibinfo  {journal} {Rev.
  Mod. Phys.}\ }\textbf {\bibinfo {volume} {82}},\ \bibinfo {pages} {1959}
  (\bibinfo {year} {2010})}\BibitemShut {NoStop}%
\bibitem [{\citenamefont {Chen}\ \emph {et~al.}(2014)\citenamefont {Chen},
  \citenamefont {Niu},\ and\ \citenamefont {MacDonald}}]{Chen_Anomalous_2014}%
  \BibitemOpen
  \bibfield  {author} {\bibinfo {author} {\bibfnamefont {H.}~\bibnamefont
  {Chen}}, \bibinfo {author} {\bibfnamefont {Q.}~\bibnamefont {Niu}}, \ and\
  \bibinfo {author} {\bibfnamefont {A.}~\bibnamefont {MacDonald}},\ }\href
  {\doibase 10.1103/PhysRevLett.112.017205} {\bibfield  {journal} {\bibinfo
  {journal} {Phys. Rev. Lett.}\ }\textbf {\bibinfo {volume} {112}},\ \bibinfo
  {pages} {017205} (\bibinfo {year} {2014})}\BibitemShut {NoStop}%
\bibitem [{\citenamefont {Ikhlas}\ \emph {et~al.}(2017)\citenamefont {Ikhlas},
  \citenamefont {Tomita}, \citenamefont {Koretsune}, \citenamefont {Suzuki},
  \citenamefont {Daisuke}, \citenamefont {Arita}, \citenamefont {Otani},\ and\
  \citenamefont {Nakatsuji}}]{Ikhlas_Large_2017}%
  \BibitemOpen
  \bibfield  {author} {\bibinfo {author} {\bibfnamefont {M.}~\bibnamefont
  {Ikhlas}}, \bibinfo {author} {\bibfnamefont {T.}~\bibnamefont {Tomita}},
  \bibinfo {author} {\bibfnamefont {T.}~\bibnamefont {Koretsune}}, \bibinfo
  {author} {\bibfnamefont {M.}~\bibnamefont {Suzuki}}, \bibinfo {author}
  {\bibfnamefont {N.}~\bibnamefont {Daisuke}}, \bibinfo {author} {\bibfnamefont
  {R.}~\bibnamefont {Arita}}, \bibinfo {author} {\bibfnamefont
  {Y.}~\bibnamefont {Otani}}, \ and\ \bibinfo {author} {\bibfnamefont
  {S.}~\bibnamefont {Nakatsuji}},\ }\href {\doibase 10.1038/nphys4181}
  {\bibfield  {journal} {\bibinfo  {journal} {Nat. Phys.}\ }\textbf {\bibinfo
  {volume} {13}},\ \bibinfo {pages} {1085} (\bibinfo {year}
  {2017})}\BibitemShut {NoStop}%
\bibitem [{\citenamefont {Zhou}\ \emph {et~al.}(2016)\citenamefont {Zhou},
  \citenamefont {Liang}, \citenamefont {Weng}, \citenamefont {Chen},
  \citenamefont {Yao}, \citenamefont {Chen}, \citenamefont {Dong},\ and\
  \citenamefont {Guo}}]{Zhou_Predicted_2016}%
  \BibitemOpen
  \bibfield  {author} {\bibinfo {author} {\bibfnamefont {J.}~\bibnamefont
  {Zhou}}, \bibinfo {author} {\bibfnamefont {Q.}~\bibnamefont {Liang}},
  \bibinfo {author} {\bibfnamefont {H.}~\bibnamefont {Weng}}, \bibinfo {author}
  {\bibfnamefont {Y.}~\bibnamefont {Chen}}, \bibinfo {author} {\bibfnamefont
  {S.}~\bibnamefont {Yao}}, \bibinfo {author} {\bibfnamefont {Y.}~\bibnamefont
  {Chen}}, \bibinfo {author} {\bibfnamefont {J.}~\bibnamefont {Dong}}, \ and\
  \bibinfo {author} {\bibfnamefont {G.}~\bibnamefont {Guo}},\ }\href {\doibase
  10.1103/PhysRevLett.116.256601} {\bibfield  {journal} {\bibinfo  {journal}
  {Phys. Rev. Lett.}\ }\textbf {\bibinfo {volume} {116}},\ \bibinfo {pages}
  {256601} (\bibinfo {year} {2016})}\BibitemShut {NoStop}%
\bibitem [{Note1()}]{Note1}%
  \BibitemOpen
  \bibinfo {note} {Several mechanisms are known to generate NCM structures:
  Dzyaloshinskii-Moriya interaction, which is a particular form of SOI arising
  in systems without inversion center, or the electron-electron Coulomb
  interaction in a system with trigonal or hexagonal lattice symmetry, where it
  effectively creates frustrated exchange interactions among spins\cite
  {Batista_Frustration_2016, Okubo_Multiple_2012}.}\BibitemShut {Stop}%
\bibitem [{\citenamefont {Nagaosa}\ and\ \citenamefont
  {Tokura}(2013)}]{Nagaosa_Topological_2013}%
  \BibitemOpen
  \bibfield  {author} {\bibinfo {author} {\bibfnamefont {N.}~\bibnamefont
  {Nagaosa}}\ and\ \bibinfo {author} {\bibfnamefont {Y.}~\bibnamefont
  {Tokura}},\ }\href@noop {} {\bibfield  {journal} {\bibinfo  {journal} {Nat.
  Nanotech.}\ }\textbf {\bibinfo {volume} {8}},\ \bibinfo {pages} {899}
  (\bibinfo {year} {2013})}\BibitemShut {NoStop}%
\bibitem [{\citenamefont {Shiomi}\ \emph {et~al.}(2013)\citenamefont {Shiomi},
  \citenamefont {Kanazawa}, \citenamefont {Shibata}, \citenamefont {Onose},
  \citenamefont {Tokura}, \citenamefont {Shiomi}, \citenamefont {Kanazawa},
  \citenamefont {Shibata}, \citenamefont {Onose},\ and\ \citenamefont
  {Tokura}}]{Shiomi_Topological_2013}%
  \BibitemOpen
  \bibfield  {author} {\bibinfo {author} {\bibfnamefont {Y.}~\bibnamefont
  {Shiomi}}, \bibinfo {author} {\bibfnamefont {N.}~\bibnamefont {Kanazawa}},
  \bibinfo {author} {\bibfnamefont {K.}~\bibnamefont {Shibata}}, \bibinfo
  {author} {\bibfnamefont {Y.}~\bibnamefont {Onose}}, \bibinfo {author}
  {\bibfnamefont {Y.}~\bibnamefont {Tokura}}, \bibinfo {author} {\bibfnamefont
  {Y.}~\bibnamefont {Shiomi}}, \bibinfo {author} {\bibfnamefont
  {N.}~\bibnamefont {Kanazawa}}, \bibinfo {author} {\bibfnamefont
  {K.}~\bibnamefont {Shibata}}, \bibinfo {author} {\bibfnamefont
  {Y.}~\bibnamefont {Onose}}, \ and\ \bibinfo {author} {\bibfnamefont
  {Y.}~\bibnamefont {Tokura}},\ }\href {\doibase 10.1103/PhysRevB.88.064409}
  {\bibfield  {journal} {\bibinfo  {journal} {Phys. Rev. B}\ }\textbf {\bibinfo
  {volume} {88}},\ \bibinfo {pages} {064409} (\bibinfo {year}
  {2013})}\BibitemShut {NoStop}%
\bibitem [{\citenamefont {Hirokane}\ \emph {et~al.}(2016)\citenamefont
  {Hirokane}, \citenamefont {Tomioka}, \citenamefont {Imai}, \citenamefont
  {Maeda},\ and\ \citenamefont {Onose}}]{Hirokane_Longitudinal_2016}%
  \BibitemOpen
  \bibfield  {author} {\bibinfo {author} {\bibfnamefont {Y.}~\bibnamefont
  {Hirokane}}, \bibinfo {author} {\bibfnamefont {Y.}~\bibnamefont {Tomioka}},
  \bibinfo {author} {\bibfnamefont {Y.}~\bibnamefont {Imai}}, \bibinfo {author}
  {\bibfnamefont {A.}~\bibnamefont {Maeda}}, \ and\ \bibinfo {author}
  {\bibfnamefont {Y.}~\bibnamefont {Onose}},\ }\href {\doibase
  10.1103/PhysRevB.93.014436} {\bibfield  {journal} {\bibinfo  {journal} {Phys.
  Rev. B}\ }\textbf {\bibinfo {volume} {93}},\ \bibinfo {pages} {014436}
  (\bibinfo {year} {2016})}\BibitemShut {NoStop}%
\bibitem [{\citenamefont {Mizuta}\ and\ \citenamefont
  {Ishii}(2016)}]{Mizuta_Large_2016}%
  \BibitemOpen
  \bibfield  {author} {\bibinfo {author} {\bibfnamefont {Y.}~\bibnamefont
  {Mizuta}}\ and\ \bibinfo {author} {\bibfnamefont {F.}~\bibnamefont {Ishii}},\
  }\href {\doibase 10.1038/srep28076} {\bibfield  {journal} {\bibinfo
  {journal} {Sci. Rep.}\ }\textbf {\bibinfo {volume} {6}},\ \bibinfo {pages}
  {28076} (\bibinfo {year} {2016})}\BibitemShut {NoStop}%
\bibitem [{\citenamefont {Hamamoto}\ \emph {et~al.}(2015)\citenamefont
  {Hamamoto}, \citenamefont {Ezawa},\ and\ \citenamefont
  {Nagaosa}}]{Hamamoto_Quantized_2015}%
  \BibitemOpen
  \bibfield  {author} {\bibinfo {author} {\bibfnamefont {K.}~\bibnamefont
  {Hamamoto}}, \bibinfo {author} {\bibfnamefont {M.}~\bibnamefont {Ezawa}}, \
  and\ \bibinfo {author} {\bibfnamefont {N.}~\bibnamefont {Nagaosa}},\ }\href
  {\doibase 10.1103/PhysRevB.92.115417} {\bibfield  {journal} {\bibinfo
  {journal} {Phys. Rev. B}\ }\textbf {\bibinfo {volume} {92}},\ \bibinfo
  {pages} {115417} (\bibinfo {year} {2015})}\BibitemShut {NoStop}%
\bibitem [{\citenamefont {Ohuchi}\ \emph {et~al.}(2015)\citenamefont {Ohuchi},
  \citenamefont {Kozuka}, \citenamefont {Uchida}, \citenamefont {Ueno},
  \citenamefont {Tsukazaki},\ and\ \citenamefont
  {Kawasaki}}]{Ohuchi_Topological_2015}%
  \BibitemOpen
  \bibfield  {author} {\bibinfo {author} {\bibfnamefont {Y.}~\bibnamefont
  {Ohuchi}}, \bibinfo {author} {\bibfnamefont {Y.}~\bibnamefont {Kozuka}},
  \bibinfo {author} {\bibfnamefont {M.}~\bibnamefont {Uchida}}, \bibinfo
  {author} {\bibfnamefont {K.}~\bibnamefont {Ueno}}, \bibinfo {author}
  {\bibfnamefont {A.}~\bibnamefont {Tsukazaki}}, \ and\ \bibinfo {author}
  {\bibfnamefont {M.}~\bibnamefont {Kawasaki}},\ }\href {\doibase
  10.1103/PhysRevB.91.245115} {\bibfield  {journal} {\bibinfo  {journal} {Phys.
  Rev. B}\ }\textbf {\bibinfo {volume} {91}} (\bibinfo {year} {2015}),\
  10.1103/PhysRevB.91.245115}\BibitemShut {NoStop}%
\bibitem [{Note2()}]{Note2}%
  \BibitemOpen
  \bibinfo {note} {A class of topologically non-trivial phase, characterized by
  Chern number $\protect \mathcal {C}$, which is a topological invariant and
  determines a quantized conductivity of AHE as $\sigma _{xy} = \protect
  \mathcal {C}(e^2/h)$.}\BibitemShut {Stop}%
\bibitem [{Note3()}]{Note3}%
  \BibitemOpen
  \bibinfo {note} {See for example, the description between Eq.(1) and (2) of
  Ref.\protect \rev@citealpnum {Mizuta_Large_2016}}\BibitemShut {NoStop}%
\bibitem [{\citenamefont {Wachter}(1972)}]{Wachter_Optical_1972}%
  \BibitemOpen
  \bibfield  {author} {\bibinfo {author} {\bibfnamefont {P.}~\bibnamefont
  {Wachter}},\ }\href@noop {} {\bibfield  {journal} {\bibinfo  {journal} {C R C
  Crit. Rev. Solid State}\ }\textbf {\bibinfo {volume} {3}},\ \bibinfo {pages}
  {189} (\bibinfo {year} {1972})}\BibitemShut {NoStop}%
\bibitem [{Note4()}]{Note4}%
  \BibitemOpen
  \bibinfo {note} {\protect \leavevmode {\protect \color {black}The general
  description with $\protect \bm {\Omega }(\protect \bm {k})$ in $\protect \bm
  {k}$-space can be translated into real space picture, in the approximation of
  strong Hund's coupling, i.e. frozen spin, where the ``spin-less" electrons
  feel ``real" magnetic field $B_{\protect \rm spin}$\cite
  {Hamamoto_Quantized_2015, Gbel_Unconventional_2017}.}}\BibitemShut {Stop}%
\bibitem [{\citenamefont {Schoenes}\ and\ \citenamefont
  {B}(1974)}]{Schoenes_Exchange_1974}%
  \BibitemOpen
  \bibfield  {author} {\bibinfo {author} {\bibfnamefont {J.}~\bibnamefont
  {Schoenes}}\ and\ \bibinfo {author} {\bibfnamefont {W.~P.}\ \bibnamefont
  {B}},\ }\href {\doibase 10.1103/PhysRevB.9.3097} {\bibfield  {journal}
  {\bibinfo  {journal} {Phys. Rev. B}\ }\textbf {\bibinfo {volume} {9}},\
  \bibinfo {pages} {3097} (\bibinfo {year} {1974})}\BibitemShut {NoStop}%
\bibitem [{\citenamefont {Ozaki}\ \emph {et~al.}()\citenamefont {Ozaki} \emph
  {et~al.}}]{openmx}%
  \BibitemOpen
  \bibfield  {author} {\bibinfo {author} {\bibfnamefont {T.}~\bibnamefont
  {Ozaki}} \emph {et~al.},\ }\href@noop {} {}\bibinfo {note}
  {\url{http://www.openmx-square.org/}}\BibitemShut {NoStop}%
\bibitem [{\citenamefont {Mostofi}\ \emph {et~al.}(2014)\citenamefont
  {Mostofi}, \citenamefont {Yates}, \citenamefont {Pizzi}, \citenamefont
  {Lee},\ and\ \citenamefont {Souza}}]{Mostofi_In_2014}%
  \BibitemOpen
  \bibfield  {author} {\bibinfo {author} {\bibfnamefont {A.}~\bibnamefont
  {Mostofi}}, \bibinfo {author} {\bibfnamefont {J.}~\bibnamefont {Yates}},
  \bibinfo {author} {\bibfnamefont {G.}~\bibnamefont {Pizzi}}, \bibinfo
  {author} {\bibfnamefont {Y.}~\bibnamefont {Lee}}, \ and\ \bibinfo {author}
  {\bibfnamefont {I.}~\bibnamefont {Souza}},\ }\href@noop {} {\bibfield
  {journal} {\bibinfo  {journal} {Comput. Phys. Commun.}\ }\textbf {\bibinfo
  {volume} {185}},\ \bibinfo {pages} {2309} (\bibinfo {year}
  {2014})}\BibitemShut {NoStop}%
\bibitem [{\citenamefont {Pizzi}\ \emph {et~al.}(2014)\citenamefont {Pizzi},
  \citenamefont {Volja}, \citenamefont {Kozinsky}, \citenamefont {Fornari},\
  and\ \citenamefont {Marzari}}]{Pizzi_BoltzWann_2014}%
  \BibitemOpen
  \bibfield  {author} {\bibinfo {author} {\bibfnamefont {G.}~\bibnamefont
  {Pizzi}}, \bibinfo {author} {\bibfnamefont {D.}~\bibnamefont {Volja}},
  \bibinfo {author} {\bibfnamefont {B.}~\bibnamefont {Kozinsky}}, \bibinfo
  {author} {\bibfnamefont {M.}~\bibnamefont {Fornari}}, \ and\ \bibinfo
  {author} {\bibfnamefont {N.}~\bibnamefont {Marzari}},\ }\href {\doibase
  10.1016/j.cpc.2013.09.015} {\bibfield  {journal} {\bibinfo  {journal}
  {Comput. Phys. Commun.}\ }\textbf {\bibinfo {volume} {185}},\ \bibinfo
  {pages} {422} (\bibinfo {year} {2014})}\BibitemShut {NoStop}%
\bibitem [{Note5()}]{Note5}%
  \BibitemOpen
  \bibinfo {note} {See for example Eq.(2) of Ref.\protect \rev@citealpnum
  {Mizuta_Large_2016} for the evaluated mathematical expressions.}\BibitemShut
  {Stop}%
\bibitem [{\citenamefont {Miyazaki}\ \emph {et~al.}(2009)\citenamefont
  {Miyazaki}, \citenamefont {Ito}, \citenamefont {Im}, \citenamefont {Yagi},
  \citenamefont {Kato}, \citenamefont {Soda},\ and\ \citenamefont
  {Kimura}}]{Miyazaki_Direct_2009}%
  \BibitemOpen
  \bibfield  {author} {\bibinfo {author} {\bibfnamefont {H.}~\bibnamefont
  {Miyazaki}}, \bibinfo {author} {\bibfnamefont {T.}~\bibnamefont {Ito}},
  \bibinfo {author} {\bibfnamefont {H.}~\bibnamefont {Im}}, \bibinfo {author}
  {\bibfnamefont {S.}~\bibnamefont {Yagi}}, \bibinfo {author} {\bibfnamefont
  {M.}~\bibnamefont {Kato}}, \bibinfo {author} {\bibfnamefont {K.}~\bibnamefont
  {Soda}}, \ and\ \bibinfo {author} {\bibfnamefont {S.}~\bibnamefont
  {Kimura}},\ }\href {\doibase 10.1103/PhysRevLett.102.227203} {\bibfield
  {journal} {\bibinfo  {journal} {Phys. Rev. Lett.}\ }\textbf {\bibinfo
  {volume} {102}},\ \bibinfo {pages} {227203} (\bibinfo {year}
  {2009})}\BibitemShut {NoStop}%
\bibitem [{\citenamefont {Arai}\ and\ \citenamefont
  {Hatsugai}(2009)}]{Arai_Quantum_2009}%
  \BibitemOpen
  \bibfield  {author} {\bibinfo {author} {\bibfnamefont {M.}~\bibnamefont
  {Arai}}\ and\ \bibinfo {author} {\bibfnamefont {Y.}~\bibnamefont
  {Hatsugai}},\ }\href {\doibase 10.1103/PhysRevB.79.075429} {\bibfield
  {journal} {\bibinfo  {journal} {Phys. Rev. B}\ }\textbf {\bibinfo {volume}
  {79}},\ \bibinfo {pages} {075429} (\bibinfo {year} {2009})}\BibitemShut
  {NoStop}%
\bibitem [{\citenamefont {G{\"o}bel}\ \emph {et~al.}(2017)\citenamefont
  {G{\"o}bel}, \citenamefont {Mook}, \citenamefont {Henk},\ and\ \citenamefont
  {Mertig}}]{Gbel_Unconventional_2017}%
  \BibitemOpen
  \bibfield  {author} {\bibinfo {author} {\bibfnamefont {B.}~\bibnamefont
  {G{\"o}bel}}, \bibinfo {author} {\bibfnamefont {A.}~\bibnamefont {Mook}},
  \bibinfo {author} {\bibfnamefont {J.}~\bibnamefont {Henk}}, \ and\ \bibinfo
  {author} {\bibfnamefont {I.}~\bibnamefont {Mertig}},\ }\href@noop {}
  {\bibfield  {journal} {\bibinfo  {journal} {Phys. Rev. B}\ }\textbf {\bibinfo
  {volume} {95}},\ \bibinfo {pages} {094413} (\bibinfo {year}
  {2017})}\BibitemShut {NoStop}%
\bibitem [{\citenamefont {Garrity}\ and\ \citenamefont
  {Vanderbilt}(2014)}]{Garrity_Chern_2014}%
  \BibitemOpen
  \bibfield  {author} {\bibinfo {author} {\bibfnamefont {K.~F.}\ \bibnamefont
  {Garrity}}\ and\ \bibinfo {author} {\bibfnamefont {D.}~\bibnamefont
  {Vanderbilt}},\ }\href {\doibase 10.1103/PhysRevB.90.121103} {\bibfield
  {journal} {\bibinfo  {journal} {Phys. Rev. B}\ }\textbf {\bibinfo {volume}
  {90}},\ \bibinfo {pages} {121103} (\bibinfo {year} {2014})}\BibitemShut
  {NoStop}%
\bibitem [{\citenamefont {Miyazaki}\ \emph {et~al.}(2010)\citenamefont
  {Miyazaki}, \citenamefont {Im}, \citenamefont {Terashima}, \citenamefont
  {Yagi}, \citenamefont {Kato}, \citenamefont {Soda}, \citenamefont {Ito},\
  and\ \citenamefont {Kimura}}]{Miyazaki_La_2010}%
  \BibitemOpen
  \bibfield  {author} {\bibinfo {author} {\bibfnamefont {H.}~\bibnamefont
  {Miyazaki}}, \bibinfo {author} {\bibfnamefont {H.}~\bibnamefont {Im}},
  \bibinfo {author} {\bibfnamefont {K.}~\bibnamefont {Terashima}}, \bibinfo
  {author} {\bibfnamefont {S.}~\bibnamefont {Yagi}}, \bibinfo {author}
  {\bibfnamefont {M.}~\bibnamefont {Kato}}, \bibinfo {author} {\bibfnamefont
  {K.}~\bibnamefont {Soda}}, \bibinfo {author} {\bibfnamefont {T.}~\bibnamefont
  {Ito}}, \ and\ \bibinfo {author} {\bibfnamefont {S.}~\bibnamefont {Kimura}},\
  }\href {\doibase 10.1063/1.3416911} {\bibfield  {journal} {\bibinfo
  {journal} {Appl. Phys. Lett.}\ }\textbf {\bibinfo {volume} {96}},\ \bibinfo
  {pages} {232503} (\bibinfo {year} {2010})}\BibitemShut {NoStop}%
\bibitem [{Note6()}]{Note6}%
  \BibitemOpen
  \bibinfo {note} {\protect \leavevmode {\protect \color {black}For the
  semiclassical picture of band conduction to be valid, $E_{\protect \rm F}\tau
  \gg 1$ is required, and therefore we should limit ourselves to $E_{\protect
  \rm F} \gg 50\ (5)$meV for $\tau =10\ (100)$ fs. To discuss as large $N$ as
  possible but safely enough, we specifically discuss $\mu \sim \pm 30$
  meV.}}\BibitemShut {Stop}%
\bibitem [{Note7()}]{Note7}%
  \BibitemOpen
  \bibinfo {note} {\protect \leavevmode {\protect \color {black}This is still
  not enough for wide applications, on which $|N| \sim 50\ \mu {\protect \rm
  V/K}$ will have a significant impact.\cite
  {Sakuraba_Potential_2016}}}\BibitemShut {NoStop}%
\bibitem [{\citenamefont {Pu}\ \emph {et~al.}(2008)\citenamefont {Pu},
  \citenamefont {Chiba}, \citenamefont {Matsukura}, \citenamefont {Ohno},\ and\
  \citenamefont {Shi}}]{Pu_Mott_2008}%
  \BibitemOpen
  \bibfield  {author} {\bibinfo {author} {\bibfnamefont {Y.}~\bibnamefont
  {Pu}}, \bibinfo {author} {\bibfnamefont {D.}~\bibnamefont {Chiba}}, \bibinfo
  {author} {\bibfnamefont {F.}~\bibnamefont {Matsukura}}, \bibinfo {author}
  {\bibfnamefont {H.}~\bibnamefont {Ohno}}, \ and\ \bibinfo {author}
  {\bibfnamefont {J.}~\bibnamefont {Shi}},\ }\href@noop {} {\bibfield
  {journal} {\bibinfo  {journal} {Phys. Rev. Lett.}\ }\textbf {\bibinfo
  {volume} {101}},\ \bibinfo {pages} {117208} (\bibinfo {year}
  {2008})}\BibitemShut {NoStop}%
\bibitem [{Note8()}]{Note8}%
  \BibitemOpen
  \bibinfo {note} {\protect \leavevmode {\protect \color {black}The band
  filling corresponds to $E_F$ close to a van Hove singularity(VHS). Drastical
  change of Chern numbers in the vicinity of VHS has already been observed\cite
  {Arai_Quantum_2009, Gbel_Unconventional_2017}, which received a novel
  interpretation recently\cite {Naumis_Topological_2016}.}}\BibitemShut {Stop}%
\bibitem [{Note9()}]{Note9}%
  \BibitemOpen
  \bibinfo {note} {\protect \leavevmode {\protect \color {black}The $|r_H| \sim
  0.2$ is one order of magnitude larger than the typical largest value obtained
  in ``bad metal"(localization) regime. See for example the discussion in
  Supporting Online Material of Ref.\protect \rev@citealpnum
  {Lee_Dissipationless_2004}}.}\BibitemShut {Stop}%
\bibitem [{\citenamefont {Lee}\ \emph {et~al.}(2004{\natexlab{a}})\citenamefont
  {Lee}, \citenamefont {Watauchi}, \citenamefont {Miller}, \citenamefont
  {Cava},\ and\ \citenamefont {Ong}}]{Lee_Anomalous_2004}%
  \BibitemOpen
  \bibfield  {author} {\bibinfo {author} {\bibfnamefont {W.~L.}\ \bibnamefont
  {Lee}}, \bibinfo {author} {\bibfnamefont {S.}~\bibnamefont {Watauchi}},
  \bibinfo {author} {\bibfnamefont {V.}~\bibnamefont {Miller}}, \bibinfo
  {author} {\bibfnamefont {R.}~\bibnamefont {Cava}}, \ and\ \bibinfo {author}
  {\bibfnamefont {N.}~\bibnamefont {Ong}},\ }\href {\doibase
  10.1103/PhysRevLett.93.226601} {\bibfield  {journal} {\bibinfo  {journal}
  {Phys. Rev. Lett.}\ }\textbf {\bibinfo {volume} {93}},\ \bibinfo {pages}
  {226601} (\bibinfo {year} {2004}{\natexlab{a}})}\BibitemShut {NoStop}%
\bibitem [{\citenamefont {Xie}\ \emph {et~al.}(2014)\citenamefont {Xie},
  \citenamefont {Bell}, \citenamefont {Kim}, \citenamefont {Inoue},
  \citenamefont {Hikita},\ and\ \citenamefont {Hwang}}]{Xie_Quantum_2014}%
  \BibitemOpen
  \bibfield  {author} {\bibinfo {author} {\bibfnamefont {Y.}~\bibnamefont
  {Xie}}, \bibinfo {author} {\bibfnamefont {C.}~\bibnamefont {Bell}}, \bibinfo
  {author} {\bibfnamefont {M.}~\bibnamefont {Kim}}, \bibinfo {author}
  {\bibfnamefont {H.}~\bibnamefont {Inoue}}, \bibinfo {author} {\bibfnamefont
  {Y.}~\bibnamefont {Hikita}}, \ and\ \bibinfo {author} {\bibfnamefont
  {H.}~\bibnamefont {Hwang}},\ }\href@noop {} {\bibfield  {journal} {\bibinfo
  {journal} {Sol. Stat. Commun.}\ }\textbf {\bibinfo {volume} {197}},\ \bibinfo
  {pages} {25} (\bibinfo {year} {2014})}\BibitemShut {NoStop}%
\bibitem [{\citenamefont {Brinkman}\ \emph {et~al.}(2007)\citenamefont
  {Brinkman}, \citenamefont {Huijben}, \citenamefont {Zalk}, \citenamefont
  {Huijben}, \citenamefont {Zeitler}, \citenamefont {Maan}, \citenamefont
  {Wiel}, \citenamefont {Rijnders}, \citenamefont {Blank},\ and\ \citenamefont
  {Hilgenkamp}}]{Brinkman_Magnetic_2007}%
  \BibitemOpen
  \bibfield  {author} {\bibinfo {author} {\bibfnamefont {A.}~\bibnamefont
  {Brinkman}}, \bibinfo {author} {\bibfnamefont {M.}~\bibnamefont {Huijben}},
  \bibinfo {author} {\bibfnamefont {V.~M.}\ \bibnamefont {Zalk}}, \bibinfo
  {author} {\bibfnamefont {J.}~\bibnamefont {Huijben}}, \bibinfo {author}
  {\bibfnamefont {U.}~\bibnamefont {Zeitler}}, \bibinfo {author} {\bibfnamefont
  {J.}~\bibnamefont {Maan}}, \bibinfo {author} {\bibfnamefont {W.}~\bibnamefont
  {Wiel}}, \bibinfo {author} {\bibfnamefont {G.}~\bibnamefont {Rijnders}},
  \bibinfo {author} {\bibfnamefont {D.~H.}\ \bibnamefont {Blank}}, \ and\
  \bibinfo {author} {\bibfnamefont {H.}~\bibnamefont {Hilgenkamp}},\
  }\href@noop {} {\bibfield  {journal} {\bibinfo  {journal} {Nat. Mater.}\
  }\textbf {\bibinfo {volume} {6}},\ \bibinfo {pages} {493} (\bibinfo {year}
  {2007})}\BibitemShut {NoStop}%
\bibitem [{\citenamefont {Wang}\ \emph {et~al.}(2009)\citenamefont {Wang},
  \citenamefont {Niranjan}, \citenamefont {Burton},\ and\ \citenamefont
  {B}}]{Wang_Prediction_2009}%
  \BibitemOpen
  \bibfield  {author} {\bibinfo {author} {\bibfnamefont {Y.}~\bibnamefont
  {Wang}}, \bibinfo {author} {\bibfnamefont {M.}~\bibnamefont {Niranjan}},
  \bibinfo {author} {\bibfnamefont {J.}~\bibnamefont {Burton}}, \ and\ \bibinfo
  {author} {\bibfnamefont {A.~J.}\ \bibnamefont {B}},\ }\href {\doibase
  10.1103/PhysRevB.79.212408} {\bibfield  {journal} {\bibinfo  {journal} {Phys.
  Rev. B}\ }\textbf {\bibinfo {volume} {79}},\ \bibinfo {pages} {212408}
  (\bibinfo {year} {2009})}\BibitemShut {NoStop}%
\bibitem [{\citenamefont {Ahadi}\ \emph {et~al.}(2017)\citenamefont {Ahadi},
  \citenamefont {Galletti},\ and\ \citenamefont
  {Stemmer}}]{Ahadi_Evidence_2017}%
  \BibitemOpen
  \bibfield  {author} {\bibinfo {author} {\bibfnamefont {K.}~\bibnamefont
  {Ahadi}}, \bibinfo {author} {\bibfnamefont {L.}~\bibnamefont {Galletti}}, \
  and\ \bibinfo {author} {\bibfnamefont {S.}~\bibnamefont {Stemmer}},\ }\href
  {\doibase 10.1063/1.4997498} {\bibfield  {journal} {\bibinfo  {journal}
  {Appl. Phys. Lett.}\ }\textbf {\bibinfo {volume} {111}},\ \bibinfo {pages}
  {172403} (\bibinfo {year} {2017})}\BibitemShut {NoStop}%
\bibitem [{\citenamefont {Batista}\ \emph {et~al.}(2016)\citenamefont
  {Batista}, \citenamefont {Lin}, \citenamefont {Hayami},\ and\ \citenamefont
  {Kamiya}}]{Batista_Frustration_2016}%
  \BibitemOpen
  \bibfield  {author} {\bibinfo {author} {\bibfnamefont {C.~D.}\ \bibnamefont
  {Batista}}, \bibinfo {author} {\bibfnamefont {S.}~\bibnamefont {Lin}},
  \bibinfo {author} {\bibfnamefont {S.}~\bibnamefont {Hayami}}, \ and\ \bibinfo
  {author} {\bibfnamefont {Y.}~\bibnamefont {Kamiya}},\ }\href {\doibase
  10.1088/0034-4885/79/8/084504} {\bibfield  {journal} {\bibinfo  {journal}
  {Rep. Prog. Phys.}\ }\textbf {\bibinfo {volume} {79}},\ \bibinfo {pages}
  {084504} (\bibinfo {year} {2016})}\BibitemShut {NoStop}%
\bibitem [{\citenamefont {Okubo}\ \emph {et~al.}(2012)\citenamefont {Okubo},
  \citenamefont {Chung},\ and\ \citenamefont {Kawamura}}]{Okubo_Multiple_2012}%
  \BibitemOpen
  \bibfield  {author} {\bibinfo {author} {\bibfnamefont {T.}~\bibnamefont
  {Okubo}}, \bibinfo {author} {\bibfnamefont {S.}~\bibnamefont {Chung}}, \ and\
  \bibinfo {author} {\bibfnamefont {H.}~\bibnamefont {Kawamura}},\ }\href
  {\doibase 10.1103/PhysRevLett.108.017206} {\bibfield  {journal} {\bibinfo
  {journal} {Phys. Rev. Lett.}\ }\textbf {\bibinfo {volume} {108}},\ \bibinfo
  {pages} {017206} (\bibinfo {year} {2012})}\BibitemShut {NoStop}%
\bibitem [{\citenamefont {Naumis}(2016)}]{Naumis_Topological_2016}%
  \BibitemOpen
  \bibfield  {author} {\bibinfo {author} {\bibfnamefont {G.~G.}\ \bibnamefont
  {Naumis}},\ }\href {\doibase 10.1016/j.physleta.2016.03.022} {\bibfield
  {journal} {\bibinfo  {journal} {Phys. Lett. A}\ }\textbf {\bibinfo {volume}
  {380}},\ \bibinfo {pages} {1772} (\bibinfo {year} {2016})}\BibitemShut
  {NoStop}%
\bibitem [{\citenamefont {Lee}\ \emph {et~al.}(2004{\natexlab{b}})\citenamefont
  {Lee}, \citenamefont {Watauchi}, \citenamefont {Miller}, \citenamefont
  {Cava},\ and\ \citenamefont {Ong}}]{Lee_Dissipationless_2004}%
  \BibitemOpen
  \bibfield  {author} {\bibinfo {author} {\bibfnamefont {W.~L.}\ \bibnamefont
  {Lee}}, \bibinfo {author} {\bibfnamefont {S.}~\bibnamefont {Watauchi}},
  \bibinfo {author} {\bibfnamefont {V.}~\bibnamefont {Miller}}, \bibinfo
  {author} {\bibfnamefont {R.}~\bibnamefont {Cava}}, \ and\ \bibinfo {author}
  {\bibfnamefont {N.}~\bibnamefont {Ong}},\ }\href {\doibase
  10.1126/science.1094383} {\bibfield  {journal} {\bibinfo  {journal}
  {Science}\ }\textbf {\bibinfo {volume} {303}},\ \bibinfo {pages} {1647}
  (\bibinfo {year} {2004}{\natexlab{b}})}\BibitemShut {NoStop}%
\end{thebibliography}%
\end{document}